# EQUS - helping to see formulae

Chris Roast
Department of Computing and Computing & Communication Research Centre
Sheffield Hallam University, Sheffield, United Kingdom
c.r.roast@shu.ac.uk

**ABSTRACT**

*Visualisation is often presented as a means of simplifying information and helping people understand complex data. In this paper we describe the design development and evaluation of an interactive visualisation for spreadsheet formula (EQUS). The work is justified on the grounds that these are widely used tools for significant numerical processing and modelling, yet the formula developed can be easily misunderstood. The development process was one of iterative refinement engaging an initial target audience of mid-teen learners, involving re-design and formative evaluation. The resulting visualisation techniques have been found to be broadly relevant to spreadsheet users beyond the initial target audience. EQUS has since been developed as fully integrated plug-in for MS Excel.*

## 1 INTRODUCTION

Visualisation is often presented as a means of simplifying information and helping people understand complex data. In this paper we describe the design and development of interactive visualisation designed to help understand spreadsheets. Our premise is that, spreadsheets are a traditional, common and accessible ICT tool that is often used to perform any number of numeric activities. Widely used in work and education (Chambers et al. (2012)), at school level and in higher education, the spreadsheet is a core generic tool for understanding in many numerate subjects. Their responsiveness means users quickly become embedded in 'solutions' in an information infrastructure that is intrinsically easy to mismanage.

Despite the difficulties that arise from embracing a spreadsheet infrastructure, the scale of their user population and widespread familiarity make them a legitimate area of study. Despite the spreadsheet being a familiar tool for general purpose computation, with significant longevity, it is widely established that they are difficult to manage (Panko and Sprague (1998); Hendry and Green (1994)). Disappointingly, the research behind the work reported occasionally encountered educational contexts in which tutors themselves did little to encourage effective spreadsheet use. It is interesting to note, that skills based training with spreadsheets involves very little modelling and numerical work, but focuses more on duplicating prescribed models and ensuring that presentation and formatting is focused upon.

Research into addressing issues of spreadsheet quality has motivated very many enhancements. This includes additional features to ensure they are more transparent as well as to encourage more discipline in their use. For some examples, see: Burnett et al. (2001, 2002); Hendry and Green (1994); Hermans and Dig (2014); Panko and Sprague (1998); Sajaniemi (2000); Ayalew (2009). Specifically, with regard to the complexity of inter-cell referencing, the understanding of formulae has been found to be particularly demanding, with evidence that business and governmental spreadsheets tend to avoid the use of many functions and function nesting (as with Sajaniemi (2000)).

### 1.1 Motivation

The starting point of the work was a recognition that initial numerical modelling with spreadsheets benefits from working with easily manipulated and refined formulae. Once a model formula is stable and not being experimented with, it is possible that the formula will be embodied in surrogate structures, such as sub-computations in separate cells. However, to maintain flexibility when exploring solutions there is a benefit to not having to commit to surrogate structures.

Previous work has proposed ways of presenting and visualising spreadsheets, see: Saariluoma and Sajaniemi (1994); Igarashi et al. (1998); Ballinger et al. (2003); Burnett et al. (2001). However, these works on the whole consider the wider structure of spreadsheets, and the dependencies between cells. None appear to have addressed the fact that the formula's language is computationally powerful but contracted onto a single line. It is this complexity of language presentation that can impair its effective use, especially when experimenting with formulae.

This observation motivated consideration of spreadsheet formulae as, in effect, small programs expressed in an inappropriate environment. One potential improvement would be to have a clearer visual language that graphically represents spreadsheet formulae. To this end we took a lead from work on program comprehension and visual programming. We chose to develop a visualisation to take advantage of human perceptual ability to recognise patterns and associations - and support "visual thinking". Graphical representations, such as flowcharts and pictorial representations of data structures have long been used to support the understanding of programs and their underlying processes (Myers 1986). However, it is of interest to note that in visual computational language, empirical evidence of their compelling and appealing character is limited (Sorva et al. (2013).

**Motivating Potential Users** Initially, tutors and learners were targeted as end users of our tool, with a focus upon numerate subjects and disciplines. With numeracy at the core of most effective spreadsheet use, we motivated end user engagement initially with the question:

$$\text{What does} 2 + 3 \times 4 \text{ equal?}$$

This was effective as a means of engaging the user community, since it was sufficiently simple to allow an open discussion of why it is useful to understand formula structure. If the user has no expectation of the result, they'll accept 14 (or 20) without worrying about what they meant by the formula. By contrast, if the user has an expectation of the result and output does not match it, they'll need support in understanding whether their formula or their expectation is wrong.

Although this case is trivial the same arguments are just as applicable for more complex formulae. Reflecting this point, our focus was to treat the visualisation as a re-presentation of a formula that would at least help disabuse users of misconceptions about the formula.

## 2 DESIGN AND DEVELOPMENT

In this section we outline our approach to design and development of EQUS. This was an iterative process informed by various factors: knowledge of visual programming research, professional graphic design and user feedback. The overall structure and phases of development were largely governed by features of technical development.

### 2.1 Designing a visual language

The visualisations were initially developed on paper to allow tutors and learners to explore and provide rapid feedback on which visual characteristics are appropriate and of value. Initially good visual design practice was followed, informed by learning scenarios and educational uses of spreadsheets (e.g. see: Gretton and Challis (2008)). The principles for the initial design where:

- Evidencing structure. Within a given formula, the syntactic structure is core to comprehending meaning.

- Visual mapping. The ease of mapping between the formula and visualisation. Clearly, if this mapping is complex for a learner, the visualisation may be of little value.
- Evidencing categories. Within a given formula, being able to recognise the different categories of tokens and structures.
- Evidencing abstractions. There are various abstractions apparent in the way formulae are used. For example, the same sub-expression appearing in a number of places in a single formula. A simple example would be the formula for a quadratic, such as, =A1*X1*X1+B1*X1+C1. The repeated use of X1 is important for understanding what is expressed.
- Evidencing computation. In contrast to abstractions, there is the value of evidencing the specific values used in determining the resulting value of a formula.
- Visual simplicity and scalability. Although not easily defined, this principle discourages apparently empty space, redundant arcs or overlapping lines or structures. In view of our motivation, this point is most relevant for complex formulae.

Clearly these principles can guide design decisions, though they can conflict with each other. As such they provide a basis for interpreting design changes and user feedback.

**2.2 Interaction Design**

In addition to providing visualisations, the manner in which users should interact and work with them was explored. In the context of this work, the objective was to enhance the use of spreadsheets and not undermine existing users skills. Within the applied educational context, departing from established materials and tutor knowledge was considered in appropriate. Hence, the interaction design was developed with aim of adding value without undermining existing skills. This perspective informed how the interaction between the tool and the spreadsheet should operate. The design thus focused upon the tool to re-present formulae responsively with the following rationale:

- The visualisation for the formula matches the formula in the current cell. If no cell is selected or it is empty, the visualisation is blank.
- The visualisation should be interactively integrated, in the sense of following the conventions of other supplementary interactive tools in spreadsheets.

These rationale in principle were to minimised the operational cognitive for spreadsheets users when working with the new visualisation.

**2.3 Development**

As described above, the design and development followed a user centred design approach, with frequent formative evaluation by end users. This progressed with both a professional visual designer engaged as well as technical developers for prototype ideas. Throughout the process each engagement with end users generated issues and questions about EQUS. These were assessed and potential tool enhancements were identified. Enhancements were judged based on relevance, taking into account prior research, and also in terms of technical feasibility.

**Paper prototyping**

To help explore the wide range of potential visualisations and help ensure users engaged in early design decisions, paper prototyping of visualisations was initially employed. This was useful in also establishing existing skill and knowledge levels of potential users. All the designs tended to use a flow visualisation aiming to convey data, dependencies and computational information. Key design decisions focused around whether to visualise structure purely in terms of operators alone (taking up less visual space) or to visualise the value computed by each intermediate operator. Despite the latter being more visually complex, it was preferred by users and could also be mapped to the principle of evidencing computation (Leitao and Roast,2014).

**Operational prototypes**

Once key visual design features had been agree upon, prototypes for dynamically generating visualisation were built. The first of these were developed in python and loosely integrated with MS Excel. This initial proof of concept helped tackle some of the initial technical questions of how the visualisation should operate. The second prototype required stronger integration with the spreadsheet and thus was developed in Javascript and to inter-operate with GoogleSheets. This provided a sufficiently robust prototype for real-world trailing and feedback.

Feedback as a whole was positive and compelling (see table 1), and used to justify moving towards a full product. Although GoogleSheets is growing in popularity, MS Excel was judged to be the primary target for a full product version. The tool was developed as a plug-in, integrated using Visual Studio Tools for Office (VSTO).

**Product development**

During product development, user feedback gathered which informed more design details (summarised in table 2).Engaging users highlighted spreadsheet usage and uses that had not initially been identified. These provoked technical and design revisions. During this phase of development additional qualitative feedback reaffirmed the potential value of elements of this design (see table 3).Despite the lack of familiarity with the visualisations, their presence and use did not impair learner performance. In follow-on interviews all agreed that the visualisation approach had merit. Overall feedback was positive, with those interviewed seeing the potential to help "de-mystify" spreadsheets for learner populations we are targeting.

The outcome of this process resulted in the EQUS tool which is described and illustrated in the following section.

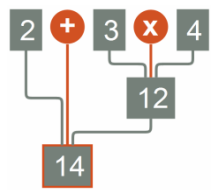

Figure 1: The formula: =2+3*4 visualised by EQUS

Table 1: Summary of initial evaluation studies and outcomes

| **Study** | **Population and context** | **Outcome** |
|---|---|---|
| Paper based study with spreadsheet comprehension questions | 44 work-based learners studying Electrical engineering, Engineering and Maths at NVQ level 3. | Visualisation showed a positive effect. Average score was 55% with a visualisation, and 51% without. (Not significant) |
| Interactive prototype based study and qualitative interviews | 14 full-time learners were given a spreadsheet "refresher" and then completed 37 spreadsheet formula questions | Visualisation conditions showed a positive effect over the no visualisation condition. Average score 73% with a visualisation, and 67% without. (Not significant) |

| Interviews and demonstration with experts. | Three STEM educators, three STEM education researchers and five support staff | Positive feedback with specific formative feedback. |
|---|---|---|

Table 2: Summary of follow-on evaluation studies and outcomes

| Study | Population and context | Outcome |
|---|---|---|
| Tutor lead usage by learners in Science education classes: GCSE and A level. Linked with the analysis of data gathered in earlier lesson | Number of learners est. 75. | Positive tutor feedback on the value of the tool for enhancing learner familiarity with BODMAS, and the physics equations that they have been learning. |
| Controlled study with participants attempting a range of given spreadsheets tasks and formulae based questions, working both with and without EQUS. | 42 higher education students (about 50 minutes of spreadsheet tasks, with and without EQUS) | Highly positive feedback. 70% would re-use EQUS, 91% would recommend it to others, 64% felt their understanding had improved. |

Table 3: Selective evaluation quotes

| Source | Quotation |
|---|---|
| School child | "It would help me a lot with other formulas" |
| School child | "You can see the values and how they are worked out, that's great." |
| Secondary level Physics tutor | "I'd draw a diagram like this on the whiteboard, but it would take a while and I might get it wrong." |
| Functional skill tutor | "Absolutely brilliant when it comes to more complicated formulas for our learners. With regards to the IF statement, I particularly like the way it checks the condition and identifies whether it is TRUE or FALSE. Additionally really good for formulas of non-adjacent cells." |
| Tutor | "I am sure that it could add value to the teaching of mathematics." |
| Tutor | "It would help anyone willing to learn about spreadsheets" |
| Maths education researcher | "It will be very useful to many students to have a product that enables a better conceptual understanding of the equation format. There is a clear need for such a tool to be suitable for the many students who do not have high levels of mathematical skills and yet use mathematical symbolism every day in their studies. This will include students from Chemistry, Business, Economics, Psychology, Geography and many more." |
| Trainee STEM tutor | "I struggle a lot with spreadsheets and find it hard to understand them. Seeing the spreadsheet visualisation prototype made it clearer to understand the formulas and feel that if I had chance to use a programme of that kind I would have a greater understanding and be able to pick up the skills I require much quicker. I feel that this product could help people like myself that struggle with spreadsheets." |

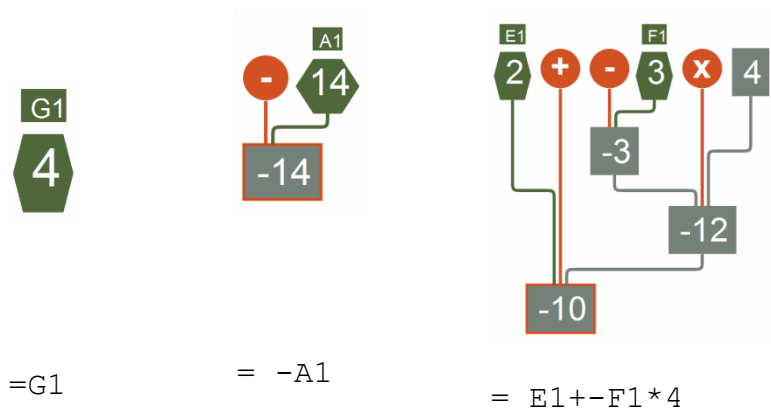

```
=G1              = -A1

                                 = E1+-F1*4
```

Figure 2: Various examples of formulae involving cell references visualised by EQUS

## 3 EQUS

Here we describe the EQUS tool introducing features that proved to be useful for a range of end users, illustrating various cases.

### 3.1 Basic formula

As mentioned in the introduction, some of the most basic formulae have presented problems for some of our initial end users. In keeping with a data flow model and the desire to see intermediate results, the visualisation for 2 + 3 × 4 is shown in figure 1.This has been found to be of most use in dynamic settings, where simply being able to quickly see what is happening, and try alternatives, has been useful. In some cases the fact that a spreadsheet is the basis for the visualisation has been of secondary interest.

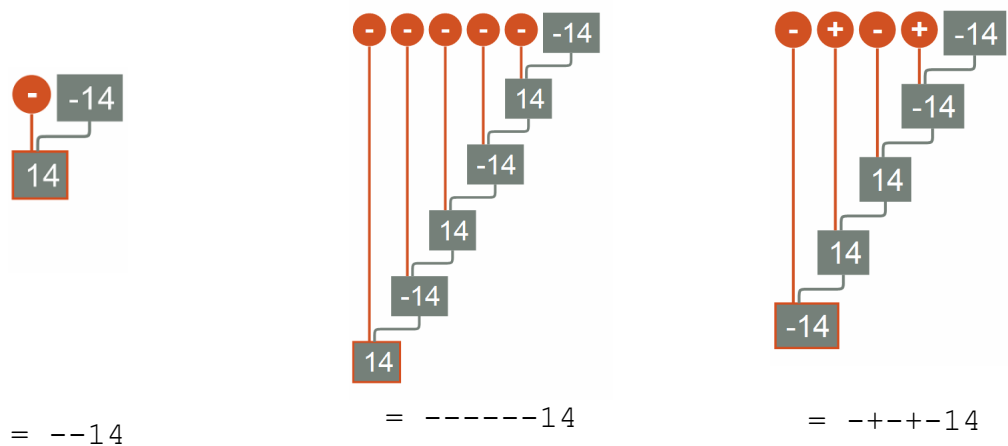

```
= --14           = ------14           = -+-+-14
```

Figure 3: Examples of prefix operators being visualised in EQUS

Cell References Spreadsheet utility primarily comes from the use of cell references for arbitrary values. Within the EQUS visualisation this is shown by differentiating the shape cell values used in a formula and an additional label for the cell reference. Cell referencing examples are shown in figure 2.The compositional character of mathematical expressions is intrinsic to their power. Interesting cases can easily be explored with the EQUS visualisations responding to any numerical expression entered into the spreadsheet. Some compelling illustrations of this include unary pre-fix operators - and +. While learning, such unary operators are, not surprising, confused with the same infix binary operators, subtract (-) and add (+) . Figure 3 shows some of the examples of repeated uses of unary operators that help illustrate their structural character and their function.

Building upon this, examples of negations interacting are useful, especially when working with non-associate operators. The importance of non-associative operators is that the order of operands makes a difference to the result, despite there being little to distinguish associative and non-associative operators by appearance. Figure 4 shows a variety of EQUS visualisations for formulae involving negation and non-associative binary operators.

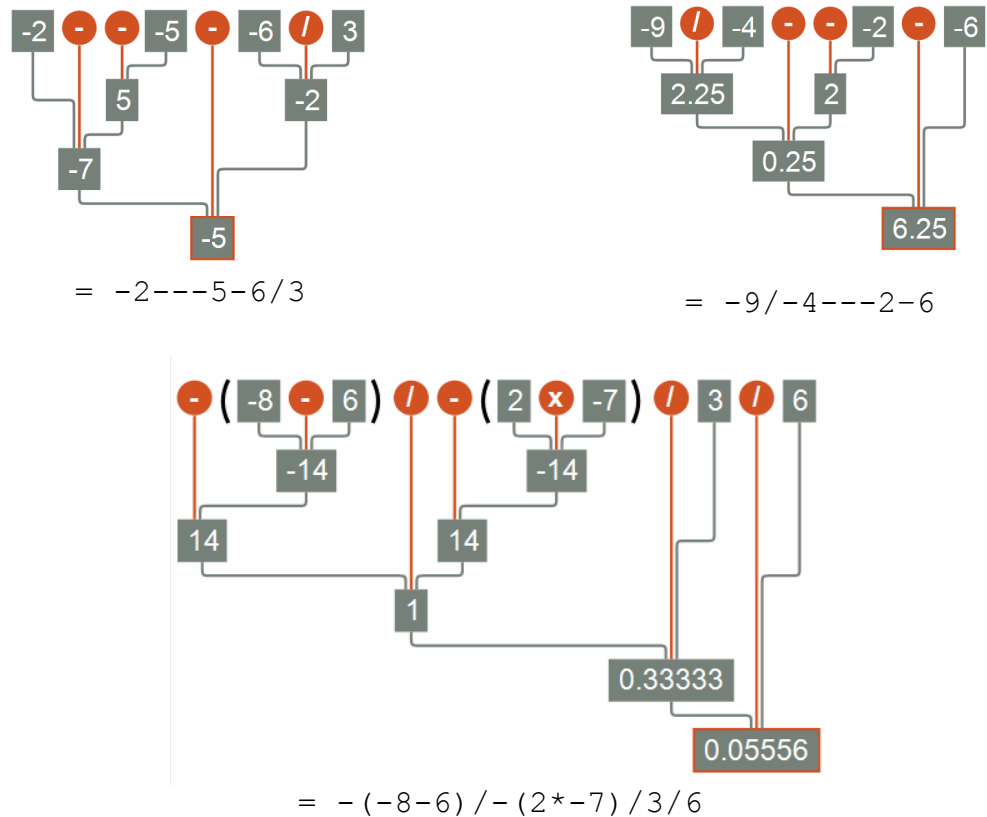

Figure 4: A number of operator based formulae visualised in EQUS

### 3.2 Functions

With spreadsheet functions EQUS visually encodes them keeping to the principles of ensuring that the formula maps to the visualisation. Figure 5 illustrates a number of standard functions combined and nested. The benefit of the EQUS visualisation is that the intermediate results are evident and therefore helping users comprehend why a result is what was intended. For real number results and intermediate values, users' expectations are likely to be estimates. However, with more discrete domains specific values are core to verifying how data is being processed. One good example of this can be found with the Boolean functions that spreadsheets support. In figure 6, the EQUS visualisation shows various Boolean expressions being computed.

### 3.3 Errors

Operator, functions and cell linking in combination give spreadsheets their power. One consequence of this is also the possibility of formulae and results being invalid and/or generating errors. During EQUS development, erroneous formulae were largely avoided. However, it become obvious as the work progressed that errors were part and parcel of spreadsheet usage - both a cause of annoyance for users but also a formative mechanism when developing numeric skills.

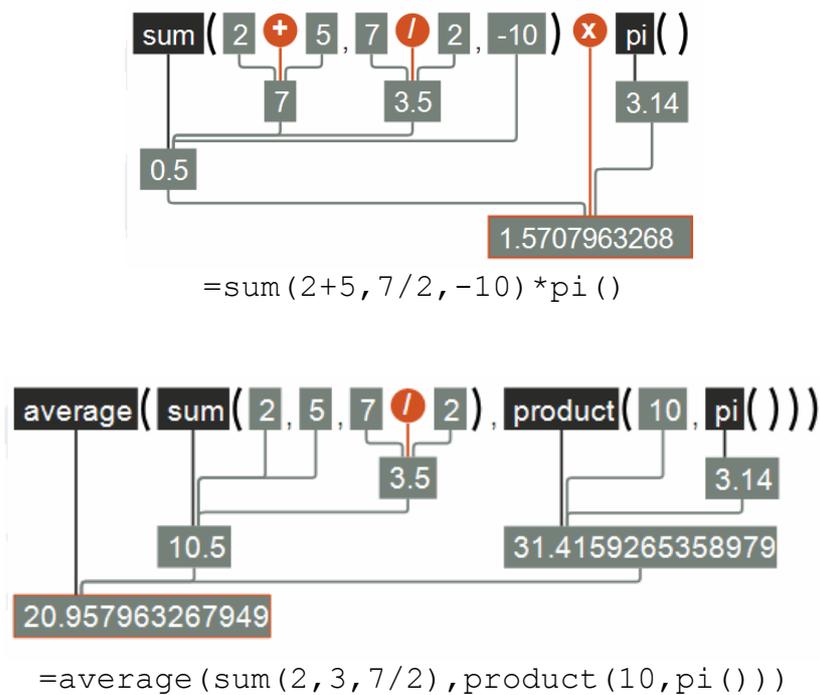

Figure 5: Example formulae with common spreadsheet functions shown in EQUS

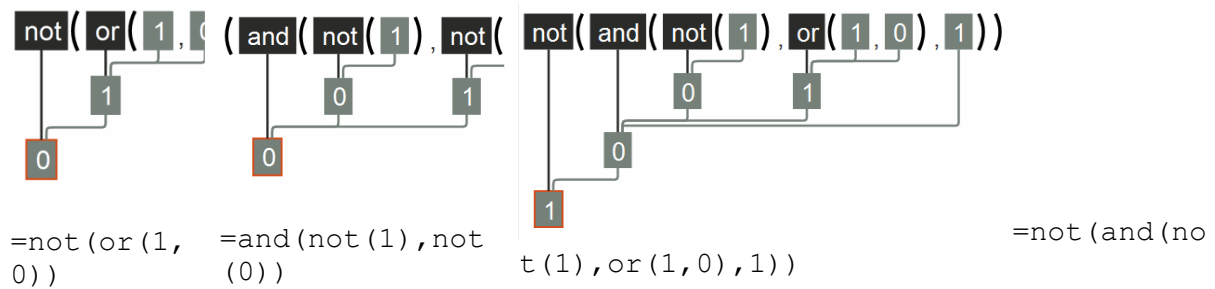

Figure 6: Example Boolean formulae visualised in EQUS

For standard spreadsheets two types of error should be distinguished. First, an invalided expression error arises from an input that simply is not a formula. In MS Excel invalided expressions are prevented from being entered, and a valid alternative expression is proposed. Second, an erroneous result arises from a syntactically valid formula that cannot be evaluated to produce a normal result. In terms of EQUS's technical design as a plug-in, there is no means of accessing invalid expressions with MS Excel, and thus, in keeping with our interaction design principles, there is no visualisation treatment of them. However, although erroneous results are not "normal" results for a formula, within MS Excel such errors are treated as to the normal domain of values that functions and operations work over. To illustrate this point, a trigonometric function, such as, `tan` would be mathematically characterised as operating over the domain of real numbers (**R**), as:

$$\texttt{tan} : \mathbf{R} \rightarrow \mathbf{R}$$

The same function when digitally computed would be best characterised as:

$$\texttt{tan} : \mathbf{R_D} \rightarrow \mathbf{R_D}$$

This characterisation, reflect the imperfect representation of the real numbers in digital technology using $\mathbf{R_D}$, as opposed to **R**. However, the same function tan in MS Excel would be characterised as operating over the same domain extended and some additional distinct error values:

$$\texttt{tan} : (\mathbf{R_D} \cup \mathbf{\mathit{E}}) \rightarrow (\mathbf{R_D} \cup \mathbf{\mathit{E}})$$

Here, **_E_** represents error-type values that extend the conventional domain of `tan` — in MS Excel there are eight such defined error-type values. The same domain extension is applied to all functions and operators in MS Excel. So for functions, such as `sqrt` and `trunk` their characterisation would be:

$$\texttt{sqrt} : (\mathbf{R_D} \cup \mathbf{\mathit{E}}) \rightarrow (\mathbf{R_D} \cup \mathbf{\mathit{E}})$$

$$\texttt{trunc} : (\mathbf{R_D} \cup \mathbf{\mathit{E}}) \times (\mathbf{R_D} \cup \mathbf{\mathit{E}}) \rightarrow (\mathbf{R_D} \cup \mathbf{\mathit{E}})$$

And so on, for all functions and operations.

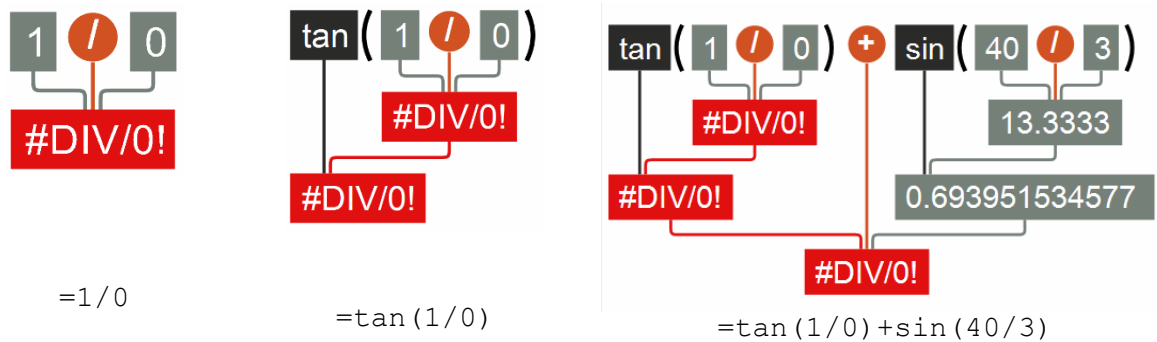

Figure 7: Three progressively complex examples of erroneous outputs visualised in EQUS

The simplest error to demonstrate this extended domain is that which arises from a division by zero. (Formally, dividing a number by zero has no sensible numeric value and thus it constitutes an erroneous result — looking like a valid expression, but with no proper result.) In the case of MS Excel, dividing by zero `=1/0` results in an error "#DIV/0!", and thus = `tan(1/0)` reflects this by also giving same error, as does = `tan(1/0)+sin(40/3)`, and so on. Any expression will compute a result when an error is present, and on most occasions the result is also the same error. Feedback from potential users about erroneous results and MS Excel's treatment of them led to the refinement of EQUS, so as to operationally treat error-type values as any other legitimate value. However, their status as "abnormal" results was made distinctive in the resulting visualisation. The examples from above are illustrated in figure 7.

A valuable result of the visualisation is that it shows how an error propagates through a computation. Thus it is possible to see where it originates and what parts of the expression are in effect error free.

**4 CONCLUSIONS**

We have reported the iterative development of the EQUS visualisation tool in terms of the design and development process, and details of its operation and resulting design. Further development of EQUS related products is envisaged in specific domains, including early education when BODMAS is being introduced, and more specialist spreadsheet integration with business needs. Additional functionality in many areas can be easily envisaged, although the merit and impact of specific suggestions requires careful analysis, so as to not over complicate a tool that benefits from simplicity.

In addition, there is the opportunity to develop a more sophisticated visual language, as recommended by Moody (2009), and discussed in Roast and Uruchurtu (2016). However, using a more subtle visual language makes significant demands upon the consistency and characterisation of functions and types. If the visualisation were to discriminate between types of function, the rationale for the types would have to be carefully established and validated. One basis is to help users recognise specific function types easily, however classifying functions appropriately is more complex — by popularity, by likely domain, by type of output are all viable alternatives.

This point reflects one of a number of design opportunities that EQUS has not incorporated despite compelling contexts of use. Examples of these include: presenting formulae in a manner visually consistent with their mathematical origins, for example, showing $\sqrt{A1}$ as opposed to `sqrt(A1)` and allowing users to edit a formula by manipulating the visualisation. As a spreadsheet augmentation such opportunities are hard to justify in terms of the design rationale since they risk specialising the tool and thus work against the objective of supporting spreadsheet users in general.

At a more academic level, experience with EQUS's development, refinement and evaluation helps inform research into visualisation and the exploration of how best to present what are, in effect, computations to users who are not thinking computationally. Work on visualising SQL provides an early illustration of this (see Roast et al., 2018).

## ACKNOWLEDGEMENTS

We are indebted to tutors, students and educational experts who were willing to support and participate in the evaluation and development of the tool. The technical development of the tool benefited significantly from support in part by: UK Higher Education Academy, The Small Business Research Initiative (part of Innovate UK) and Department for Business, Innovation & Skills (UK). VeryViz Ltd. is also acknowledged for supporting recent development activity.

# References

Ayalew, Y. (2009). A visualization-based approach for improving spreadsheet quality. In Proceedings of the Warm Up Workshop for ACM/IEEE ICSE 2010, WUP '09, pages 13–16, New York, NY, USA. ACM.

Ballinger, D., Biddle, R., and Noble, J. (2003). Spreadsheet structure inspection using low level access and visualisation. In Proceedings of the Fourth Australasian User Interface Conference on User Interfaces 2003 - Volume 18, AUIC '03, pages 91–94, Darlinghurst, Australia, Australia. Australian Computer Society, Inc.

Burnett, M., Atwood, J., Walpole Djang, R., Reichwein, J., Gottfried, H., and Yang, S. (2001). Forms/3: A first-order visual language to explore the boundaries of the spreadsheet paradigm. J. Funct. Program., 11(2):155–206.

Burnett, M., Sheretov, A., Ren, B., and Rothermel, G. (2002). Testing homogeneous spreadsheet grids with the "what you see is what you test" methodology. IEEE Trans. Softw. Eng., 28(6):576–594.

Chambers, C., Sommers, Z., and Scaffidi, C. (2012). A study of help requested online by spreadsheet users. J. Organ. End User Comput., 24(4):41–53.

Gretton, H. and Challis, N. (2008). Fundamental Engineering Mathematics; A Student Friendly Workbook. Woodhead Publishing.

Hendry, D. and Green, T. (1994). Creating, comprehending and explaining spreadsheets. Int. J. Hum.-Comput. Stud., 40(6):1033–1065.

Hermans, F. and Dig, D. (2014). Bumblebee: A refactoring environment for spreadsheet formulas. In Proceedings of the 22Nd ACM SIGSOFT International Symposium on Foundations of Software Engineering, FSE 2014, pages 747–750, New York, NY, USA. ACM.

Igarashi, T., Mackinlay, J. D., Chang, B.-W., and Zellweger, P. T. (1998). Fluid visualization of spreadsheet structures. In Proceedings of the IEEE Symposium on Visual Languages, VL '98, pages 118–, Washington, DC, USA. IEEE Computer Society.

Leitao, R. and Roast, C. (2014) Developing visualisations for spreadsheet formulae: towards increasing the accessibility of science, technology, engineering and maths subjects. In: 9th Workshop on Mathematical User Interfaces, Coimbra, Portugal, 10 July 2014.

Moody, D. (2009). The "physics" of notations: Toward a scientific basis for constructing visual notations in software engineering. IEEE Trans. Softw. Eng., 35(6):756–779.

Myers, B. A. (1986). Visual programming, programming by example, and program visualization: A taxonomy. SIGCHI Bull., 17(4):59–66.

Panko, R. R. and Sprague, Jr., R. H. (1998). Hitting the wall: Errors in developing and code inspecting a 'simple' spreadsheet model. Decis. Support Syst., 22(4):337–353.

Roast, C. and Uruchurtu, E. (2016) Reflecting on the Physics of Notations applied to a visualisation case study. MexIHC'16, September 21 - 23, Colima, Mexico. DOI: 10.1145/2967175.2967383.

Roast, C., Patterson, D, Hardman, V. (2018). VISUALISATION – IT IS NOT THE DATA, IT IS WHAT YOU DO WITH IT.   In:  e-Society 2018 Conference Proceedings. 231-238, IADIS. ISBN: 978-989-8533-75-3.

Saariluoma, P. and Sajaniemi, J. (1994). Transforming verbal descriptions into mathematical formulas in spreadsheet calculation. Int. J. Hum.-Comput. Stud., 41(6):915–948.

Sajaniemi, J. (2000). Modeling spreadsheet audit: A rigorous approach to automatic visualization. Journal of Visual Languages & Computing, 11(1):49 – 82.

Sorva, J., Karavirta, V., and Malmi, L. (2013). A review of generic program visualization systems for introductory programming education. Trans. Comput. Educ., 13(4):15:1–15:64.